\documentclass[12pt]{article} 
\usepackage{amsmath}
\usepackage{amssymb}
\begin{document}  
\title{\bf About the intuitive picture of a Hamiltonian for a dissipative system}
\author{ G. V. L\'opez\\ \\
\small Departamento de F\'{\i}sica de la Universidad de Guadalajara\\
\small Apartado Postal 4-137, Ad. de correos no. 4\\
\small 44421 Guadalajara, Jalisco, M\'exico}  

\maketitle

\begin{abstract}
A Hamiltonian for a one-dimensional (1-D) dissipative system is given which shows that the trajectories in the spaces ($x,\dot x$) and ($x,p$) are completely different. The trajectory in the space ($x,p$) has an unexpected contra-intuitive behavior, and a canonical transformation does not solve this behavior.
\end{abstract}

\newpage
\section{\label{I} Introduction}
Modern physics has its foundation mostly on the Lagrangian-Hamiltonian formalism 
(Messiah 1958,  and Toda 1992). For classical mechanics, the Lagrangian-Hamiltonian formulation is equivalent to Newton formulation whenever the Lagrangian (therefore, the Hamiltonian) of the system exists (Darboux 1894, and Douglas 1941). Of course, given the Lagrangian of the system, there is always (through the Euler-Lagrange equations) a second order differential equation. However, the converse is not always true (Douglas 1941), and this problem is called " The inverse problem of the Mechanics (or Calculus of Variations)."  
For conservative systems , where the total force of the system is a function on the position of the particle, ${\bf F(x)}$, a constant of motion is just the so called energy of the system, which is given by $E=mv^2/2+V({\bf x})$, where $v$ is the speed of the particle (where $v^2=v_x^2+v_y^2+v_z^2$ with $v_i=dx_i/dt$ the ith-component of the velocity), and ${\bf x}=(x,y,z)$ represents its position. The term $mv^2/2$ is the kinetic energy of the system, and the $V({\bf x})=-\int {\bf F(x)}\cdot d{\bf x}$ is the potential energy. The Lagrangian and the Hamiltonian are given by $L=mv^2/2-V({\bf x}) $ and $H=p^2/2m+V{\bf x})$, where $p=|{\bf p}|$ is the magnitude of the generalized linear momentum, ${\bf p}=\nabla_vL=m{\bf v}$. As it is also well known, for a 1-D conservative problems, the trajectories in the spaces ($x,v$) and ($x,p$) are totally equivalents, and the intuitive behavior of the particle can be seen in either space. Of course, once we have the trajectory in the ($x,p$) space, a canonical transformation can be made, $Q=Q(x,p)$ and $P=P(x,p)$, to get a completely different trajectory in the space $(Q,P$), for example, the usual angle-action transformation (Lichtenberg and Lieberman 1983, and Arnold 1978). However, it is known ( L\'opez 1996)
even for one-dimensional non conservative system, the trajectories on these spaces can be different.  In this paper a drastic  behavior of this type is shown, given  the trajectories in the spaces ($x,v$) and ($x,p$)  for a 1-D quadratic dissipative system. Firstly, a constant of motion, the Lagrangian, the generalized  linear momentum and the Hamiltonian are deduced for the dissipative system. Thus, the trajectories in the spaces ($x,v$) and ($x,p$) are given, showing this intuitiveness problem. Finally, it is shown that a canonical transformation does not help much with this problem.
\section{\label{II} Constant of motion, Lagrangian and Hamiltonian}
Consider the 1-D motion of a particle of constant mass $m$ in a dissipative medium characterized by the  fact that it produces a force on the particle given by $F(x)=-\alpha \dot x^2$ ($\dot x>0$), where $\alpha$ 
 is the dissipative constant. The Newton equation of motion for this system is given by
 $$m\ddot x=-\alpha \dot x^2\ ,\eqno(1)  $$
 and the constant of motion of the system is a function $K=K(x,v)$, with the definition $\dot x=v$,  satisfying $dK/dt=0$, or equivalently (L\'opez, 1999)
 $$v\frac{\partial K}{\partial x}-\alpha v^2\frac{\partial K}{\partial v}=0\ . \eqno(2) $$
 The general solution of this equation is $K=G(C)$ with $G$ being an arbitrary function of the characteristic curve $C$ (John 1974),
 $$C=v e^{\alpha x/m}\ .\eqno(3)$$
 Since for $\alpha\to 0$, the usual energy of a free particle is the usual constant of motion, $E=mv^2/2$, the functionality is chosen as $G(C)=mC^2/2$, that is, the constant of motion is chosen as
 $$K_{\alpha}(x,v)=\frac{1}{2}mv^2 e^{2\alpha x/m}\eqno(4)$$
 which has the right expected limit, $\lim_{\alpha\to 0}K_{\alpha}=mv^2/2$. \\Ê\\
 Given this constant of motion, the Lagrangian of the system can be calculated from the relationship (Kobussen 1979, Leubner 1981, L\'opez 1996)
 $$L=v\int\frac{K(x,v)}{v^2}dv\ ,\eqno(5)$$
which brings about the Lagrangian
$$L_{\alpha}(x,v)=\frac{1}{2}mv^2 e^{2\alpha x/m}\ .\eqno(6)$$
The generalized linear momentum is
$$p=mv e^{2\alpha x/m}\ ,\eqno(7)$$
and from the inverse relation
$$v=\frac{p}{m} e^{-2\alpha x/m}\eqno(8)$$
and the Legendre transformation, $H=vp-L$, the Hamiltonian of the system is
$$ H_{\alpha}=\frac{p^2}{2m} e^{-2\alpha x/m}\ .\eqno(9a)$$
This Hamiltonian generates the Hamiltonian equations
$$\dot x=\frac{p}{m} e^{-2\alpha x/m}\ ,\quad\quad \dot p=\frac{\alpha p^2}{m^2} e^{-2\alpha x/m}\ .\eqno(9b)$$
Note that these quantities have the right expected limits for $\alpha$ going to zero.  
Now, given the initial conditions ($x_o,v_o$), the constant of motion (4) is determined. Therefore, the trajectory of the particle defined by the relation
$$v=\sqrt{\frac{2K_{\alpha}}{m}}~e^{-\alpha x/m}\ .\eqno(10)$$
On the other hand, with these initial conditions and Eq. (7), the initial point ($x_o,p_o$) in the phase space is determined, and the trajectory in this space is defined by the relation
$$p=\sqrt{2mH_{\alpha}}~e^{\alpha x/m}\ .\eqno(11)$$
For example, if $x_o=0$ and $v_o>0$ (implying $p_o>0$), Eq. (10) tells us that the velocity goes to zero as the particle moves further away. This is obviously  the expected intuitive picture of the dynamics of the particle. However, Eq. (11) tells us that linear momentum goes to infinity as the position increases. This strange behavior is due to the expression (7), where the trivial relation between linear momentum  ($p=mv$) appearing for conservative systems does not appears for the dissipative system.  However, the solutions of the systems (1) and (9b) are the same, taking into account the relation (7) between the velocity and the linear momentum.
\section{\label{III} Canonical transformation}
The obvious question is whether or not a canonical transformation could restore the intuitive picture that is  expected for dissipation. Then, let us assume that there is a canonical transformation (Goldstein 1950)
 characterized by a generatrix function, says $F_1(x,Q)$,  such that 
 $$p=\frac{\partial F_1}{\partial x}\ ,\quad\quad P=-\frac{\partial F_1}{\partial Q}=\beta(x) e^{-\lambda Q}\ .\eqno(12)$$
where $\lambda$ is a real positive constant, and $\beta$ is an undefined function. Integrating these equations, it follows that 
$$F_1=\frac{\beta(x)}{\lambda} e^{-\lambda Q}+f(x)\ ,\eqno(13a)$$ 
and
$$p=\frac{\beta'}{\lambda} e^{-\lambda Q}+f'(x)\ ,\eqno(13b)$$
where $\beta'=d\beta/dx$, and $f(x)$ is arbitrary. This means that the variable $x$ becomes a complicated relation with respect $Q$ and $P$,
 $$x=\beta^{-1}\bigg(Pe^{\lambda Q}\biggr)\eqno(14a)$$
 and
 $$p=\frac{1}{\lambda}\beta'\biggl(\beta^{-1}(Pe^{\lambda Q})\biggr)+f'\biggl(\beta^{-1}(Pe^{\lambda Q})\biggr)\ ,\eqno(14b)$$
 where $\beta^{-1}$ is the inverse function of $\beta$. Assuming $f=constant$, it follows that
 $$P=\frac{\lambda\beta(x) p}{\beta'(x)}\eqno(15a)$$
 and
 $$Q=-\frac{1}{\lambda}\ln\left(\frac{\lambda p}{\beta'(x)}\right)\ .\eqno(15b)$$
 In this way, although Eq. (12) looks like the restoration of the intuitive behavior of the dissipative system in the space ($Q,P$), the complicated relationship between the variable still remains.
\section{\label{H} Conclusions }
For a 1-D quadratic dissipative system, it has been shown that the intuitive physical picture of a trajectory in the space ($x,p$) is lost, although the solution of the system is the same. Furthermore, a canonical transformation can not restore this intuitive picture without further complications. This has be done  within a consistent mathematical approach to find the associated Hamiltonian for a dissipative system.  It is my opinion that this behavior should be taken into account when one wants to study the quantization (Landsberg 1992) or the statistical properties of dissipative systems. In addition, Hamiltonian formalism is not a confident approach when one  wants to see  the qualitative behavior of a non conservative (dissipative) system.

\end{document}